\documentclass[runningheads]{llncs}

\usepackage{amsfonts}
\usepackage{amssymb}
\usepackage{graphicx}
\usepackage[utf8x]{inputenc}
\usepackage[ruled]{algorithm2e}
\usepackage{amsmath}
\usepackage{booktabs}

\begin{document}

\title{Machine Consciousness as Pseudoscience: The Myth of Conscious Machines}
\author{Eduardo C. Garrido-Merchán}
\date{March 2023}

\institute{Universidad Pontificia Comillas, Madrid, Spain \\
\email{ecgarrido@icade.comillas.edu}}

\maketitle 

\abstract{The hypothesis of conscious machines has been debated since the invention of the notion of artificial intelligence, powered by the assumption that the computational intelligence achieved by a system is the cause of the emergence of phenomenal consciousness in that system as an epiphenomenon or as a consequence of the behavioral or internal complexity of the system surpassing some threshold. As a consequence, a huge amount of literature exploring the possibility of machine consciousness and how to implement it on a computer has been published. Moreover, common folk psychology and transhumanism literature has fed this hypothesis with the popularity of science fiction literature, where intelligent robots are usually antropomorphized and hence given phenomenal consciousness. However, in this work, we argue how these literature lacks scientific rigour, being impossible to falsify the opposite hypothesis, and illustrate a list of arguments that show how every approach that the machine consciousness literature has published depends on philosophical assumptions that cannot be proven by the scientific method. Concretely, we also show how phenomenal consciousness is not computable, independently on the complexity of the algorithm or model, cannot be objectively measured nor quantitatively defined and it is basically a phenomenon that is subjective and internal to the observer. Given all those arguments we end the work arguing why the idea of conscious machines is nowadays a myth of transhumanism and science fiction culture.}


\section{Introduction}
Machine consciousness, also referred by the technical literature as artificial consciousness \cite{reggia2013rise}, has been debated both philosophically and technically since machines were able to display intelligent behaviour given a set of predefined instructions. A plethora of technical claims and architectures about how a machine may give rise to consciousness have been published in the recent years by a multidisciplinary community \cite{gamez2018human}. However, different communities do not even agree into a definition of what consciousness is, having a wide array of theories about the nature of consciousness \cite{seth2022theories}. To disambiguate this term, we focus on the study on the awareness of phenomenal consciousness, the perception of qualia by the observer, the self or life \cite{schwitzgebel2016phenomenal}. Having stated that, we observe that current machine consciousness theories and frameworks rely on philosophical assumptions when they claim to uncover the nature and generation of consciousness, such as for example the multiple realizability assumption \cite{cao2022multiple}, being the philosophical hypothesis that the a mental property, state, or event can be implemented by different physical properties, states, or events. Critically, those philosophical hypotheses can not be tested empirically, and consequently the technical frameworks that assert to achieve machine consciousness depend on the hypothesis that those philosophical assumptions are true, which we can not proven using the scientific method. 

In this work, we illustrate how the current machine consciousness theories, frameworks and technical implementations rely on philosophical assumptions that can not be proven using the scientific method, basically because consciousness cannot be measured externally to the observer nor guaranteed as a result of the behaviour of an artificial machine \cite{cole2004chinese}. Consequently, we believe that these studies, although speculative and useful for thought experiments, can not be catalogued as scientific, specially the technical frameworks that claim that consciousness can be generated, or emerge, as an epiphenomenon of the behaviour or the information processing of a system. As a consequence, the transhumanist myth of conscious machines currently belongs to the realm of science fiction, but can not be scientifically proven, being pseudoscience.

We think that is important that philosophies such as sentientism \cite{roelofs2023sentientism} understand that our current artificial intelligence, with large language models or autonomous agents trained with deep reinforcement learning \cite{li2017deep}, are not aware but are just statistical information processing architectures. In other words, the importance of not falling into the fallacy of attributing consciousness to a system that displays intelligent behaviour is critical \cite{garrido2022artificial} nor denying consciousness to, for example, humans that do not display intelligent behaviour such as people with general learning disability conditions \cite{gillberg2003learning}. It is also a research that is specially relevant if the metaverse soon becomes a reality, to differentiate between NPCs or avatars and people \cite{wang2022survey}. As a society, it is very important to overcome folk psychology derived from science fiction about robots possesing consciousness \cite{geraci2007robots} and to assess whether we can consider machines just information processors or autonomous moral agents to know which tasks can we delegate to machines and which ones not \cite{garrido2023computational}, as human beings are able to perceive qualia, which makes us different from machines \cite{balduzzi2009qualia}. 

In the following section we will provide a hierarchy of consciousness definitions and the one that we are going to study with a justification for itk, we will also illustrate how a science of the definition of consciousness that we hold is an oxymoron. Afterwards, we will continue with a section that compiles the theoretical arguments against the hypothesis of machine consciousness, dealing with the uncomputability of consciousness and philosophical arguments and thought experiments. We continue with a section that critiques current practical machine consciousness approaches and conclude with a discussion about the repercussions of this work and conclusions and further research lines.   

\section{Is it possible to define consciousness?}
Current surveys about machine consciousness focus mostly on what is defined as the scientific approach towards phenomenal consciousness \cite{butlin2023consciousness}, but, in our opinion, lack a philosophical perspective about the nature of the observer of phenomenal consciousness, which is the property that we are going to be focused on this research. Concretely, we want to study whether it is possible to talk about a science of the observer of phenomenal consciousness and why this is an oxymoron and hence conscious machines are just a myth of transhumanism. We begin this section with a proper definition of our view about consciousness, disambiguating it from other related concepts. We then argue the issues with a science of consciousness, having provided a definition of the concept. This contextualization will be useful to provide our arguments against machine consciousness in the following section. We summarize all the arguments that we mention in this section and the following one in Table \ref{arguments} with their corresponding identifiers. 

Different communities such as psychology or philosophy of mind do not tend to agree in a formal definition about what consciousness is but we build our definition based on the informal definition of Koch \cite{koch2018consciousness}, being consciousness everything that we experience, that partly satisfies what we mean about this concept. Concretely, we use that definition but extend it to define consciousness as everything that we experience, formally described as set of quale also denoted as qualia or qualia space, and that critically includes, apart from the qualia space, the fact of being able to experiment an awareness that is witness of a qualia space, different for every individual and being that experience private to an individual and subjective. 

Another qualitative not as satisfying definitions include being sentient and awake or being able to referring to properties of mental states such as feelings and understanding \cite{rosenthal2009concepts}. Trying to provide a formal definition of consciousness seems elusive as the concept is usually used to englobe multiple properties about our awareness and because multiple research communities have developed different research lines based on different definitions. That is why we here disambiguate the term into different definitions and then, once that are defined, give arguments about why our definition is the one that englobes all the rest of them and should be used to argue why is impossible to scientifically assert with absolute certainty that machines will arise consciousness.  

We need, in the first place, to differentiate between the pure awareness, or life, or the observer, or the self \cite{heidegger1927sein,masao1985self} between the rest of characteristics commonly attributed to consciousness. We are specially focused on the observer of a phenomenological space, also referred as qualia space by integrated information theory (IIT). In this context, the observer is aware of the qualia space, but somehow is like a witness of this space, that includes being aware of time, space and all the information processed by the brain, both computational and qualia. In this paper, when we talk about consciousness, we talk about the observer, that lies beyond what is usually referred as the explanatory gap \cite{levine1983materialism}.

We also find in the literature a distinction between phenomenal consciousness and access consciousness \cite{block1995confusion}. Commonly, life or the observer is usually confused with phenomenal consciousness \cite{schwitzgebel2016phenomenal}. In particular, from a philosophical perspective, life or the observer is a superset of phenomenal consciousness. Concretely, phenomenal consciousness is the perception of qualia by an observer, who is able to percieve a qualia or phenomenological space. We emphasize that although the observer has phenomenal consciousness, it also has other properties such as pure awareness, in the dasein sense of Being and Time \cite{heidegger1927sein} or how Zen Buddishm describes pure consciousness or mind essence. The observer would also have, as a result of brain activity, access consciousness \cite{block1995confusion} to the actions that brain activity choose to perform with the body as a result of the internal and external mental states that the observer perceives of the environment and internal representations of information.

Other different definitions of consciousness found in the literature are the narrative being, self or personal identity \cite{sturma2016self}, which is encoded in the brain through specific networks. It is important to differentiate between the observer and the narrative being as drug experiments demonstrate through reports of patients that substances like DMT (dimethyltryptamine) or Ketamine can produce what is called as ego death, the disappearance of the individual sense of self or narrative being, being temporally erased the perception of oneself as an entity that is separate from the external environment. However, the patients report that they were still aware, not about their identity but mixed with the environment. This is a consequence of decreased and modified brain activity as a result of drug consumption, also find at near death experiences \cite{martial2021losing}. Consequently, it is critical to differentiate between the neural representation of the narrative self, being physically encoded in the brain and manipulable through drugs, and pure awareness, which is our concept of consciousness.     

\begin{figure}[htb!]
  \centering
  \includegraphics[width=0.99\textwidth]{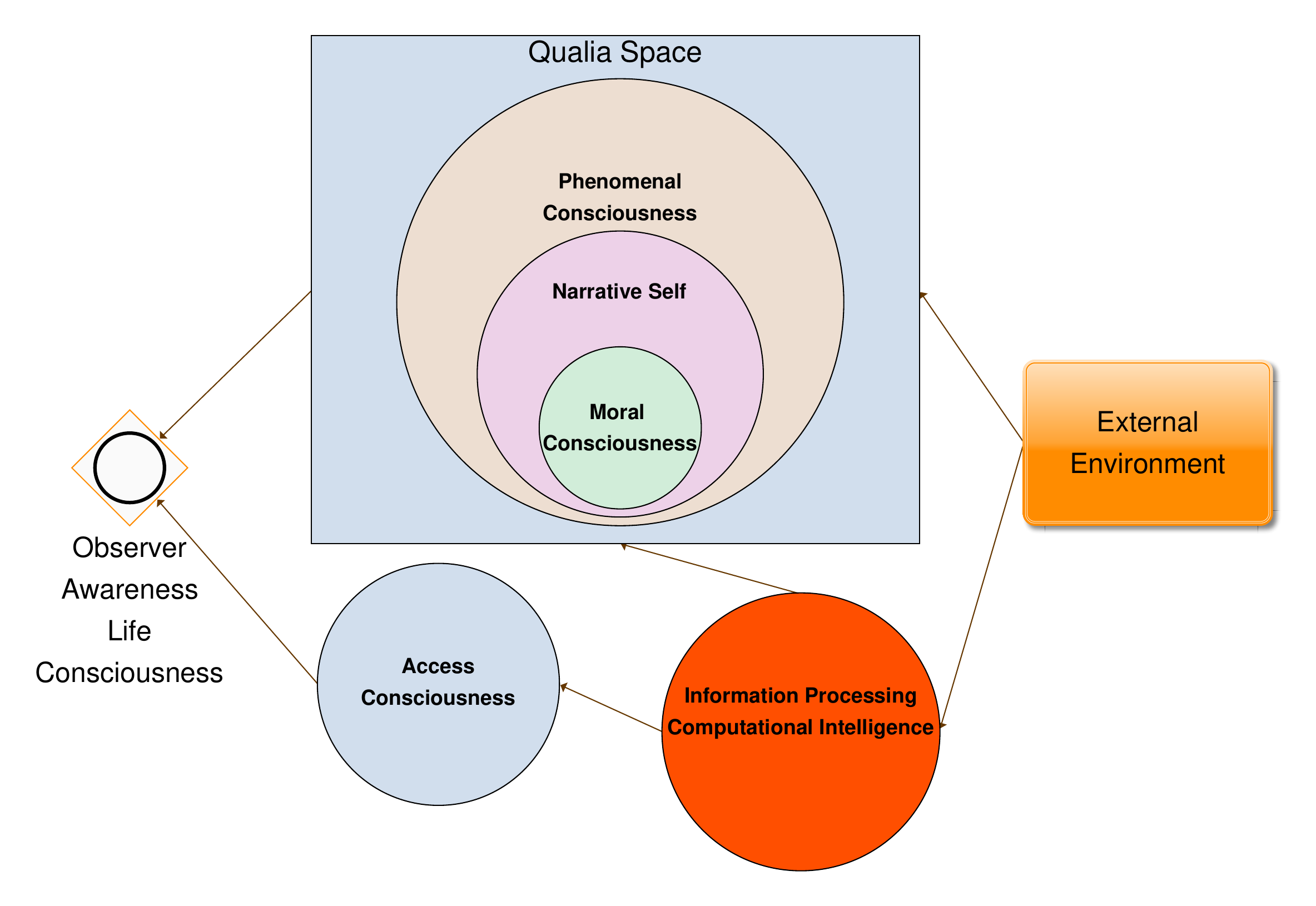}
  \caption{Consciousness categories that are illustrated in this section and input information flows to the observer. The qualia space is private and unique to every observer and is perceived as a consequence of phenomenal consciousness and the information processing and brain chemical reactions happening in the brain.}
  \label{fig:cons_flow}
\end{figure}

Finally, as a subset of phenomenal consciousness we find moral consciousness \cite{habermas1990moral}, being the perception of morality associated with an observer according to its values. Critically, the qualia moral space is a subspace of the full qualia space, which is why we argue that moral consciousness is a subset of phenomenal consciousness, perceived by the observer. Lastly, we emphasize that computational intelligence is independent of phenomenal consciousness, holding the arguments found in Garrido's work \cite{garrido2023computational}. In other words, there is no objective way to proof the assumption that by overcoming a predefined measure of complexity or a numerical threshold by a system, consciousness emerges as a result of the complexity of the system, because the complexity needs to be defined mathematically and consciousness is subjective. We will refer later to this concept as the complex threshold assumption. Having stated that, and being our object of study the observer, from now on referred as consciousness, we describe in the following subsection the issues of designing a science of consciousness. We illustrate on Figure \ref{fig:cons_flow} all the concepts seen in this section.

\subsection{A science of machine consciousness?}
Science has uncovered a universe of knowledge since the invention of the scientific method, studying mensurable quantities and retrieving empirical evidence to falsify a null hypothesis in order to accept alternative hypotheses. We need hence to falsify the opposite hypothesis to the one that we are interested on to work with the scientific method \cite{popper1963science}. However, it is impossible to falsify whether a system, agent or even a living being has phenomenal consciousness or to even measure or provide an objective definition of consciousness, we can only work with correlations and signatures extracted with mechanisms such as fRMIs and reports by human beings or simulate qualia spaces due to analogies \cite{barttfeld2015signature}. Neuroscientific studies usually talk about a science of consciousness in the access consciousness meaning provided in the previous section, which is perfectly possible as the actions performed by living beings as a function of the states perceived to maximize an expected reward are usually encoded as signatures in the human brain \cite{dehaene2021consciousness}, similarly we can encode this behaviour in machines by means of methodologies that estimate a policy of an agent such as deep reinforcement learning \cite{li2017deep}, being an autonomous agent but not a moral autonomous agent as it lacks the qualia associated to morality. 

Unfortunately, we cannot objectively measure awareness or uncover the truth about its cause, being impossible to discern whether a confounder is the responsible of the neural signatures retrieved or if it is consciousness, what is usually defined as the explanatory gap \cite{levine1983materialism,revonsuo2021explanatory}. This is a particular real example of the correlation is not causation statement found in fundamentals of statistics \cite{geer2011correlation}, which is necessary to take into account to not fall into the fallacy of the popular "fake it till make it", which is more formally known as behaviourism \cite{gehlmann2019fake}. In other words, it is not necessary to attribute awareness to a, from an external point of view, intelligent system, based on its behaviour, as consciousness and computational intelligence, or artificial general intelligence \cite{goertzel2007artificial}, do not necessarily need to be dependent nor a cause or an effect one of the other, as a confounder may perfectly explain this dependence \cite{garrido2023computational}. Consequently, we believe that as consciousness is a private to the observer and subjective phenomenon its field of study lies beyond the tools of the scientific method to different branches of philosophy. That is why, is we define pseudoscience as statements, beliefs, or practices that claim to be both scientific and factual but are incompatible with the scientific method, then, the machine consciousness practice is a pseudoscience, as we will further provide arguments for it in this section.

Lastly, we also emphazise that, following the same argument, the theory of mind object of study is what here is referred as the narrative self, and not the observer of awareness. It is important to understand that theory of mind is just a theory and cannot be scientifically proven to be true in an ontological sense \cite{leudar2009introduction}. For example, the consequences of our argument is that for the severe autism condition, which is controversial for theory of mind skeptics \cite{lacroix2023autism,frith1994autism},  we will assume that according to the theory of mind it is possible, but it cannot be proven on an ontological sense, to have a human being with severe autism without consciousness of the narrative self but the human being belonging to the severe autism condition would have awareness of phenomenal qualia and be an observer.   

To sum up, we believe that it is not possible to design a science of consciousness, if we define consciousness as life or awareness, but it is possible to define a science of access consciousness or comptuational intelligence. This is because life can not be measured directly, only relying on correlations, but without a way to ensure that the cause of consciousness does not have a confounder that cannot be detected with the scientific method, as consciousness cannot be defined, cannot be measured directly, it is a phenomenon that is private to the observer and subjective.

\section{Arguments against machine consciousness}
Having disambiguated the concept of consciousness that we are studying in the previous section and discerned why it is elusive for current science, we now use the facts described to elaborate our arguments against the hypothesis of conscious machines and why it is a myth to assert that consciousness can rise in machines, being its study a pseudoscience. The organization of the section is as follows, first we theoretically illustrate how consciousness is uncomputable and then we enumerate philosophical arguments and thought experiments about the uncomputability of consciousness. These arguments will be useful in the following section, where we will use them to critique current machine consciousness approaches. 

\subsection{The uncomputability of consciousness}
Current AI includes large language models, autonomous agents and symbolic artificial intelligence, that are software programs that run on a GPU, CPU or a similar processing unit of a computer. This high level software is translated into low assembly code and then into binary code by compilers, hardware gates and run on Von Neumann architectures. Critically, the instructions are set by the programmer and the routines are pseudo-deterministic, driven by the logic of instructions, even in the case of generating instructions recursively by the instructions set by the programmer. Specifically, large language models text generations are computed by matrix multiplications of word embeddings and neural layers and the actions performed by deep reinforcement learning agents are just matrix multiplications of the values estimated by policy learning algorithms. Critically, we can englobe all of these algorithms into the space of algorithms that a Turing machine is able to generate \cite{turing1936turing}.

Formally, a Turing machine can be represented as a 7-tuple \((Q, \Gamma, b, \Sigma, \delta, q_0, F)\), where \(Q\) is a finite set of states where the machine can be at any moment, \(\Gamma\) is a finite set that includes the tape alphabet, \(b \in \Gamma\) is the blank symbol, used to indicate empty cells on the tape, \(\Sigma \subseteq \Gamma \setminus \{b\}\) is the set of input symbols, \(\delta: Q \times \Gamma \rightarrow Q \times \Gamma \times \{L, R\}\) is the transition function, dictating the machine transitions between states left or right as a result of some input information, \(q_0 \in Q\) is the initial state or beggining of the algorithm and \(F \subseteq Q\) are the target states where the execution can be finished.

Concretely, any algorithm can be described as a finite set of deterministic rules or instructions given to the machine by the user, for example a large language model architecture and associated optimization procedure. If each instruction in an algorithm corresponds to a state transition in a Turing machine, we can build a Turing machine for any given algorithm by defining its transition function \(\delta\) to simulate each step of the algorithm. Consequently, as all components of an algorithm can be mapped to operations of a Turing machine, it follows that a Turing machine can execute any algorithm.

However, as we have stated, consciousness is not an algorithm, just awareness of mental states or qualia spaces, hence is not runnable by a Turing machine and consequently it cannot be computed. Algorithms, as purely functional and computational processes, cannot account for being witnesses of subjective experiences, as our definition of consciousness states. It is hence an observation of reality, nor a function of some inputs and output, not a mental state nor a computational state where we can make transitions on. It is just observation of the states, but not a state. But moreover, even in the case of qualia, and following the qualia argument for pure awareness, qualia are inherently non-computational, so it cannot be captured by any algorithmic description. Again, as we already hinted in previous sections, this is because algorithms process objective data and outputs based on predefined instructions, whereas qualia are inherently subjective and private to the individual. So nor phenomenal consciousness nor awareness can be encoded into a Turing machine but are independent on it, although a Turing machine can provide information to access consciousness and information for the generation of a qualia space. We do not need to involve complex concepts here as free will, as we believe that just showing how awareness is not computational as it cannot be run by a Turing machine, as it is not an algorithm, is enough to state that awareness is not computational.   

Even under the assumption that awareness may be explained not just computationally but, somehow under philosophical assumptions, described mathematically, Gödel's incompleteness theorems \cite{godel1931formal} state that any sufficiently powerful and consistent formal system cannot prove all truths about the arithmetic operations within the system; there will always be true statements that are unprovable within the system. As we do not grasp any clue due to the explanatory gap about consciousness, remaining the hard problem unsolved, we cannot prove mathematically the existence of a subjective reality that cannot be described by inference in the mathematical system. Moreover, even in the case of phenomenal consciousness, observe that if the human mind is equivalent, somehow, to a formal algorithmic system, it would be subject to Gödel's incompleteness constraints. However, human mathematicians are capable of understanding and accepting the truth of Gödel’s incompleteness theorems themselves, suggesting, again, an ability to transcend the limitations of any formal system or algorithm even in the case of phenomenal consciousness. Following this argument, Roger Penrose, who also argues that human consciousness involves non-computable processes that no algorithmic model can fully replicate or understand based on our mentioned Gödel's theorem \cite{penrose1991emperor}, proposed that consciousness arises from quantum state reductions occurring within the neural microtubules in the brain, a theory known as Orchestrated Objective Reduction (Orch-OR). However, again, we find the explanatory gap as consciousness cannot be measured, which makes these speculations valid and useful but belonging to the realm of pseudoscience as they cannot be falsified.

\subsection{Philosophical arguments and thought experiments}
After having shown the issues of asserting that a mathematical description of a technical implementation of consciousness is possible, we also elaborate some thought experiments to add theoretical evidence of our hypothesis, that consciousness cannot be measured without assuming that current correlations and brain signatures do not have an unknown confounder that we cannot access as a consequence of the explanatory gap and then, hence, machine consciousness is a pseudoscience. 

An intuitive thought experiment is the Chinese room presented by Searle \cite{cole2004chinese}. The arguments places a person inside a room with one place where it enters a chinese message and another place where an english message is the output. External observers cannot see what is happening inside the room, where the person follows a complete set of rules to map Chinese symbols to the English language. Although externally, and only based on the behaviour of the room, we could think that the person inside the room appears to understand Chinese, it actually lacks any comprehension of the language, it just follows rules as an algorithm. This also illustrates our previously defined idea that mere symbol manipulation does not necessarily generate consciousness as an epiphenomenon nor does it imply the existence of real understanding or consciousness. Daniel Dennett and others have suggested that if the Chinese Room system were embedded within a computer capable of interacting with the external world in more sophisticated ways, it might begin to be attributed some level of understanding, but this is mere speculation that relies on the multiple realizability assumption and that it fails to answer our previous arguments about the nature of consciousness, that we cannot be sure about a mathematical definition and hence about a technical implementation. In other words, there is no way to proof the multiple realizability assumption.

As we cannot be sure about qualia generation, we cannot assert that given more inputs from reality consciousness is going to emerge as an epiphenomenon. In other words, there is no objective way to proof the epiphenomenon assumption. Moreover, given Ockham's razor \cite{garrido2022artificial}, it is a more complex hypothesis to state that consciousness emerges that it does not emerge, so the hypothesis of consciousness emerging needs to be falsified. As it cannot be falsified, we cannot accept that hypothesis nor reject that it does not emerge. Consequently, it is simpler to state that no algorithm in a machine, a turing machine, makes consciousness emerge so it cannot be asserted that a bigger input information set rises consciousness, and as there is no understanding without the qualia perceived of the problem, hence, the chinese argument holds.      

Another thought experiment worth to mention is the one that is referred as the Stilwell Brain, where an enourmous number of human beings are placed in a field being organized as a Von Neumann architecture, where every human being mimicks the behaviour of a single transistor. A consequence is that all the humans simulate the behaviour of a CPU, being able to run any algorithm that can be run by a computer or computable by a Turing machine. With this experiment we get rid of the black-box feeling that the behaviour of a computer inspires. If we assume that consciousness emerges as an epiphenomenon of functional activity in a computer, it would also emerge in the Stilwell brain, which seems more complicated as it lack a physical substrate, being the only hypothesis available that consciousness is attracted from a non-physical reality by the functional activity, which as it cannot be measured nor detected, it is a more complex hypothesis that the opposite one, that consciousness does not emerge as functional activity. Hence, as we cannot falsify the null hypothesis, that it does not emerge, giving Ockham's razor, we cannot accept that it is attracted in such a way, what makes the functionalist hypothesis of consciousness emergence difficult to accept without assuming a panpsychist perspective.   

Finally, we also comment the Knowledge Argument \cite{jackson1986mary} to show why critiques against the existence of the qualia space and the observer face an important issue. The experiment involves Mary, a scientist who knows everything about the physical aspects of color but has never experienced color herself, that is, the qualia of color. Mary discovers all the inference about color given its objective information and computational intelligence, but, when Mary sees a red object for the first time she learns the qualia non-computational information, the experience red. There is no way for a computer, given previous arguments, to process qualia or report about it, as it is what consciousness experiences, and it is information, as Mary is learning something new, hence this adds evidence that supports the hypothesis that both the qualia space and the observer exists, although they cannot be mathematized nor implemented in a computer, given previous arguments.  

More thought experiments include the extension of the observer to other living beings with different qualia space such as bats that cannot be replicated in current machines \cite{nagel1980like}, World 3 argument by Popper \cite{boyd2016popper} that differentiates between physical states which is World 1, mental states which is world 2 and products of human minds which is world 3 defending that machines do not have access to world 2, which is what we have defined as the qualia space and the observer, being world 3 access consciousness. We summarize all the arguments that we have mentioned throughout this section in Table \ref{arguments} to be used in the following section to critique machine consciousness theories.

\begin{table}[h]
\centering
\caption{Arguments Against Machine Consciousness theories and technical implementations.}
\label{arguments}
\begin{tabular}{@{}lp{9cm}p{3cm}@{}}
\toprule
\textbf{Identifier} & \textbf{Argument Name} & \textbf{Type} \\ \midrule
A1 & Turing Machine computability & Mathematical \\
A2 & Godel incompleteness theorem & Mathematical \\
A3 & Ockham's razor & Statistical \\
A4 & Chinese room & Philosophical \\
A5 & Stilwell Brain & Functionalist \\
A6 & Multiple realizability assumption & Philosophical \\
A7 & World 3 (Non-locality) & Metaphysical \\
A8 & Hard problem of consciousness & Philosophical \\
A9 & Qualia space subjectivity & Phenomenological \\
A10 & Awareness indefinable formally & Philosophical \\
A11 & Epiphenomenon assumption & Philosophical \\
A12 & Complex threshold assumption & Computer science \\
A13 & Argument of Knowledge & Philosphical \\
A14 & Explanatory Gap & Phenomenological \\
A15 & Drug manipulation does not erase consciousness & Chemistry \\
A16 & Awareness is not access consciousness & Philosophical \\
A17 & Computational Intelligence depending on consciousness fallacy  & Philosophical \\
\bottomrule
\end{tabular}
\end{table}

\section{A critique on current machine consciousness approaches}
Once we have provided our view of consciousness, the issues of providing a mathematical definition or simulate consciousness in the machines and philosophical arguments against the hypothesis of conscious machines we review the machine consciousness literature describing each proposed framework, methodology, theory or implementation and use previous arguments to illustrate the issues of every proposed solution. It is important to remark that we describe consciousness theories that can be used to assert that their implementation in machines can arise consciousness. That is, when we describe a theory and provide arguments against the theory it is with respect to the implementation of the theory by some mechanism or approximation or framework into a machine or robot.

We begin with the Global Workspace Theory (GWT) \cite{baars1997theatre}, that proposes that consciousness emerges as a function of a "global workspace" in the brain, a central exchange where different cognitive processes can access and share information Global workspace theory. This shared memory accessible to another process can also be implemented on software and it is a essential tool of operative systems. As we have seen before, this theory assumes that consciousness is a function of the computational complexity of a system. Concretely, the shared memory architecture described in the theory can be easily simulated with autonomous agents sharing a shared memory, being hence a computational system that can be simulated by an algorithm, that can be run in a Turing machine. Hence, the global workspace theory needs to address to objection of the Turing machine not modelling consciousness and provide evidence that supports the claim that consciousness emerges without a measure as it is more complex and Ockham's razor applies. Notwithstanding, we believe that the theory can be accurate for the specific case of access consciousness, focused on the attention bottleneck that several components compete to occupy in conscious processing. Other computational model that attempt to perform an abstractly similar idea is the consciousness prior \cite{bengio2017consciousness}, where similar concepts are given mathematical expressions. Although useful to model access consciousness, we believe that the approach falls into Godel incompleteness theorems, as we have no way to assert that such a mathematical formulation has anything to do with awareness.

The neuroscience community has also explored the hypothesis of replicating consciousness in machines. Concretely, Dehaene differentiates two types of information-processing computations in the brain: one for selecting information for global broadcasting, defined as C1, and another one for self-monitoring these computations to generate a subjective sense of certainty or error, defined as C2 \cite{dehaene2021consciousness}. In fact, it resembles System 1 for intuitive processing and System 2 of slow processing of Kahneman \cite{kahneman2011thinking}. But it divides System 1 in C0, complex calculations and tasks that we labelled as computational intelligence that recent technological advances have enabled in machines and C1, associated with cognitive processing and access consciousness that may be simulated with autonomous policies estimated with deep reinforcement learning. However, and most critically, Dehaene agrees that cognitive C2 self-monitoring subjective tasks can not replicated by machines, it does not talk about the cause of awareness but just about correlations of neural signatures happening in the brain in conscious patients. We emphasize that neuroscience agrees that consciousness or awareness involves more than just processing information and includes subjective experiences that machines currently cannot achieve.

Another neuroscience common theory is the Attention Schema Theory, proposed by Graziano and colleagues \cite{butlin2023consciousness}, suggests that consciousness arises from how the brain manages attention, being another materialist theory of consciousness. According to this theory, the brain constructs a dynamic model of itself that explains and predicts the behavior of controlling and focusing attention. The theory posits that this model is what we experience as consciousness, allowing us to be aware of ourselves as entities with the capacity for attention. However, we believe that this theory is perfectly describing access consciousness and the narrative representation of the self but, again, does not explain how qualia arises and it basically lacks a response to the same arguments as the Global Workspace Theory. A similar materialist neuroscientific theory is the Recurrent Processing Theory \cite{eklund2012recurrent}, that basically claim that the unconscious functions of feature extraction and categorizations, as in computer science when feature engineering is performed for machine learning models, are done by the feedforward sweep, whereas conscious functions that, partly, are linked to perceptual organization are managed by recurrent (feedback, or re-entrant) cortico-cortical connections. The key idea beyond the theory is that RPT states that processing in sensory regions of the brain are sufficient for conscious experience. We find two issues with the theory. First, it is a materialist theory that, again, does not provide an answer to arguments as the Chinese room, the hard problem of consciousness or the stilwell brain. Second, as we have seen, awareness is not just qualia perception but pure observation, whose cause is not targeted by this theory. 

Other theories find the difficulty of providing a definition of consciousness as we had seen in Section 2 and to avoid the issue and critique present a circular definition, what are called as higher-order theories \cite{seth2022theories}. Concretely, and concisely, in these theories, consciousness arises when mental states are represented by higher-order mental states, due to a first-order state, a computational representation in brain signatures, being in, some ways not clearly defined, monitored or meta-represented by a relevant higher-order representation \cite{brown2019understanding}. Although the intuition accurately approximates the concept of awareness being witness of phenomenal consciousness, the higher-order representation, this theory is circular in the sense that it defines consciousness as a monitoring consciousness on a representation. Consequently, this theory is falling in the assumption that consciousness can be defined formally in an objective way, when it is a subjective concept private to the individual and hence not mensurable nor definable. 

Maybe the most famous theory about consciousness from a general point of view is Integrated Information Theory (IIT) \cite{tononi2016integrated}. Here we find a clear example of consciousness being an epiphenomenon or consequence of systems whose complexity is higher than a predefined threshold, in this case $\phi$, the complex threshold assumption whose hypothesis is more complex than just consciousness not emerging. Although being beautiful mathematically with its axiomatic definition, the theory assumes that its objective mathematical definition describes a completely subjective property as consciousness, lacking empirical validation and not adequately accounting for neurobiological properties. It also depends on heuristics like the earth moving distance that are not justified theoretically. Moreover, does not explain the explanatory gap nor the hard problem of consciousness and it also heavily relies on the multiple realizability assumption or panpsychism assumptions. Furthermore, the theory is not falsifiable, as it does not provide clear predictions that can be tested against observations or experiments. All these arguments are the reason why integrated information theory has been proclaimed as pseudoscience by a big number of researchers \cite{fleming2023integrated}. We also like to remark that the qualia space definition provided in integrated information theory that assigns to every quale a numerical dimension does not accurately model phenomenological realities. The quale of love, for example, being reduced to a simple real number is a poor representation of the plethora of feelings experienced by the aware being, better described with Bergson's intensities.  

We find also a plethora of machine consciousness implementations in robots summarized in the works of David Gamez \cite{gamez2008progress,gamez2018human} that assert to have replicated consciousness directly in machines and robots. Some examples are the CRONOS robot involving the SIMNOS simulator and the spiking neural simulator called SpikeStream, the COG robot written in Lisp, the CyberChild simulated infant controlled by a biologically-inspired neural system and other classic cognitive architectures. Gamez defines four levels of consciousness from MC1, Machines with the external behaviour associated with consciousness to MC4, Phenomenally conscious machines. All categories from MC1 to MC3 would just be machines that exhibit behaviour related to access consciousness. It is important to address that it is admitted that there is no way to objectively confirm MC4 on machines without relying on the multiple realizability assumption. Moreover, these approaches are speculative in the sense that they are theoretical frameworks hardcoded into robots trying to axiomatize consciousness mathematically rather than relying on robust empirical evidence or even correlations of brain signatures and are all falling into the behavioural assumption that we have defined as assuming consciousness in a system addressing computational intelligence behaviour, which also violates Ockham's razor. They also fall in the fallacy of addressing consciousness as access consciousness, being behavioural theories that assert that a system is conscious based on its behaviour, as the mirror test, a clear example of the Chinese Room, as everything in the agent is mere syntax but evidently lacking existence of the qualia space. 

Due to the rise of artificial intelligence several theories have used the deep reinforcement learning framework to design a behavioural theory in what is referred as consciousness inspired reinforcement learning \cite{zhao2021consciousness}. Formally, Deep reinforcement learning (DRL) uses deep learning to encode the policy of an autonomous agent learnt with reinforcement learning algorithms. The loss function used to optimize the deep neural network is design to maximize an expected cumulative reward $\mathbb{E}(r)$, defined mathematically as: 
\[
\max_\pi \mathbb{E} \left[ \sum_{t=0}^\infty \gamma^t r_t \right],
\]
where \(\pi\) is the most critical factor here as it denotes the policy, that basically at each timestep gives a function that is able to map a set of states \(S\) to a set of actions \(A\), \(r_t\) represents the reward received at every timestep \(t\), and \(\gamma\) is the discount factor $(0 < \gamma \leq 1)$, which prioritizes present over future rewards. The training period uses a simulation of the environment to explore which action $a$ is expected to achieve a maximum expected reward $\mathbb{E}(r)$ for every possible state values $\mathbf{s} \in S$, that is the autonomous agent acts as $a^* = argmax_A(\mathbf{s}, A)$, where $a^*$ is the action that maximizes $\mathbb{E}(r)$ being for every possible state $\mathbf{s}$ the distribution of states and actions the policy $\pi = p(S,A)$. The analogy is straightforward, if $S$ is mapped to the qualia space perceived by consciousness, $A$ is mapped to every action that a human being can perform at any time $t$ and the training process is represented in neural signatures, the theory can explain human behaviour. However, this theory does not take into account that qualia are non computational information, as we have seen before, so the estimated policy $\pi$ would be biased and does not explain why qualia emerges nor why does consciousness exists and assumes that consciousness is a computational function $\pi$ that maps states to actions, being a functionalist theory. We have seen that such theories do not also explain Godel Incompleteness Theorem and are algorithms, hence are Turing machines, so they are not able to model consciousness as pure awareness. This theory, however, can explain human behaviour inferred only from computational intelligence and access consciousness. 

Lastly, we would like to also affirm from a Bayesian perspective why machine consciousness must be regarded as pseudoscience. Concretely, when an object of study, such as machine consciousness in this case, has been proposed a plethora of theories where every theory seems to have the same uncertainty about its object of study, then the marginal likelihood or evidence about the hypothesis is low. In fact, there is not enough evidence to reject the method of assigning, in this case, an non informative prior, such as an uniform one, to all the theories about consciousness rising on machines. As there is no way to obtain empirical evidence to support the claim that machines are, even phenomenally conscious, then, there is no way to update the prior and obtain a posterior. Consequently, we have achieved a stable distribution where every theory has the same probability of being true or the reason why consciousness emerges. Hence, the entropy about the hypothesis of consciousness emerging is maximum, which means that we do not know anything the random variable in this study that would be the the hypothesis of conscious machines. To sum up, as there is no way to decrease the uncertainty about the existence of consciousness in machines empirically using the scientific method, our subjective beliefs can not be updated using a Bayesian perspective and hence the object of study is purely subjective, being any attempt to study it formally pseudoscience. 

We summarize consciousness theories with respect to their implementation in machines described in this section and the arguments that they lack to response to objectively assert the rise, generation or attraction of consciousness in the following table. 

\begin{table}[h]
\centering
\caption{Theories of Consciousness and arguments, whose identifier can be seen on Table 1, against their implementations on machines. We emphasize that no theory provides a solution to the hard problem of consciousness. We only include the main arguments used for every theory and exclude secondary arguments against the theory. }
\label{my-label}
\begin{tabular}{@{}p{8,2cm}p{5cm}@{}}
\textbf{Consciousness Theory implemented on machines} & \textbf{Main Arguments Against} \\ \midrule
Global Workspace Theory                         & A1, A3, A6, A8, A9, A10, A15  \\
The Consciousness Prior                         & A1, A2, A6, A8, A10, A11, A17\\
C0, C1, C2                         &  A3, A8, A9, A14, A15\\
Attention Schema Theory                         & A3, A6, A8, A9, A10, A14\\
Recurrent Processing Theory                         & A3, A6, A8, A9, A10, A14\\
Integrated Information Theory                         & A2, A3, A4, A6, A8, A9, A12, A17\\
Higher order theories                         &  A1, A2, A5, A7, A8, A10 \\
CRONOS, SIMNOS, Spikestream, COG, Cyberchild & A4, A5, A6, A8, A9, A13, A16, A17\\
Consciousness Inspired Reinforcement Learning                         &  A1, A2, A3, A5, A8, A10, A11, A12\\
\bottomrule
\end{tabular}
\end{table}

\section{Discussion}
Having reviewed the main theories of consciousness and which arguments can be used to compromise their claims about the hypothesis of rising consciousness in a machine assuming that the theories hold from a scientific and a philosophical point of view, we illustrate in this section a discussion about the nature of machine consciousness from a broader perspective.

Consciousness has been studied thorughout the history by various spiritual traditions such as Zen buddishm or Christian spirituality, with the interiority method \cite{thomas2000interiority}. The Zazen method of Buddishm gives a structured method of meditation in several phases to achieve what is called as the brilliant light, a state of pure awareness \cite{cook1999raise}. Most interestingly, tests such as fRMI empirically show how brain activity differs significantly in Buddishm monks, showing how they are able to manipulate brain signatures and hence their mental states. Drug experiments with DMT also show a manipulation of brain states with the death of the ego as extreme case. We believe that these practices can alter the qualia space, that is, the phenomenal consciousness of the observer, that is still aware but achieves to unlink pure awareness from the narrative self and the major part of the qualia space. By doing it so the observer can observe itself, the state of pure awareness. We believe that science must research these methods to grasp an understanding of pure awareness, and that these methods are completely independent on computational intelligence, hence being impossible to implement them on machines given the previous reasons. We believe, hence, that consciousness is a reality that must be researched with subjective studies, spirituality or studying reports of patients being given drugs to manipulate their consciousness, and that, being a subjective reality, machine consciousness is a discipline of pure speculation and thought experiments.  

We believe that consciousness has been accurately studied with a broader perspective in previous works such as Being and Time \cite{heidegger1927sein} where its unmodellable nature is properly described. Hence, both the observer and its qualia space cannot be mathematized nor translated into a computer without previous assumptions and consquently any attempt to further develop the literature of machine consciousness, given that in machines every model or algorithm needs to be formalized, is pseudoscience, that, however maybe useful as pure speculation or thought experiments. 

However we believe that consciousness must be researched with the tools that we have mentioned, both by neuroscience, spiritual studies and practices and by chemistry with drugs or other related substances as phenomenal consciousness affects our behavioural policy and reality, given that we are able to identify emotional mental states, react as a cause of them, like for example a person in love, and learn new non computational information given by qualia \cite{goldman1993consciousness}. Consequently, although machine consciousness seems to be a transhuman myth that cannot be solved by the scientific method, we believe that studying consciousness leaving its existence of machines apart is critical, because it is what makes us human and different from machines.

\section{Conclusions and Further Work}
Machine consciousness has been studied by a plethora of approaches, driven by the quest of making machines aware of themselves. However, the hypothesis of machine consciousness can not be falsified, being impossible to give empirical evidence incompatible with the fact of a machine not being conscious, as consciousness is a subjective phenomenon that is private to the observer and because current machine consciousness implementations and theories heavily rely on several philosophical assumptions that can not be also proven by the scientific method. 

In this paper, we have provided a list of arguments against the claim of having proven the hypothesis that machines are aware of themselves as a result of an implementation of several consciousness theories like GWT or IIT. We have also reviewed how the main machine consciousness theories can be implemented in machines and associate main arguments that apply for every machine consciousness implementation as objections to the fact that the implementation makes machines aware of themselves. We recommend, as further work, to monitorize implementations of machine consciousness theories and check whether they provide an answer to the arguments listed.   

As none of the current approaches satisfies every argument that apply, such as the hard problem of consciousness, the chinese room, modelling a non-computational reality by a Turing machine or relying on the complex threshold assumption, and the scientific method can not be used, machine consciousness is a pseudoscience and the transhuman hypothesis that machines will be conscious as a result of technological progress is a myth.  

\bibliography{main}
\bibliographystyle{acm}

\end{document}